\documentclass[12pt,preprint]{aastex}

\shorttitle{UDF Near IR Background}
\shortauthors{Thompson et al.}

\begin{document}

\title{Constraints on the Cosmic Near Infrared Background Excess from NICMOS 
Deep Field Observations}

\author{Rodger I. Thompson}
\affil{Steward Observatory, University of Arizona, Tucson, AZ 85721}
\email{rthompson@as.arizona.edu}

\author{Daniel Eisenstein}
\affil{Steward Observatory, University of Arizona,
    Tucson, AZ 85721}
\email{deisenstein@as.arizona.edu}

\author{Xiaohui Fan}
\affil{Steward Observatory, University of Arizona,
    Tucson, AZ 85721}
\email{fan@as.arizona.edu}

\author{Marcia Rieke}
\affil{Steward Observatory, University of Arizona,
    Tucson, AZ 85721}
\email{mrieke@as.arizona.edu}

\and

\author{Robert C. Kennicutt}
\affil{Institute of Astronomy, University of Cambridge, Cambridge CB3 OHA UK and Steward Observatory, 
	University of Arizona, Tucson, AZ 85721}
\email{robk@ast.cam.ac.uk}

\begin{abstract}

NICMOS observations of the resolved object fluxes in the Hubble Deep Field
North and the Hubble Ultra Deep Field are significantly below the fluxes 
attributed to a 1.4 - 1.8$\micron$ Near InfraRed Background 
Excess (NIRBE) from previous low spatial resolution NIRS measurements. Tests
placing sources in the NICMOS image with fluxes sufficient to account for
the NIRBE indicate that the NIRBE flux must be either flat on scales greater 
than $100\arcsec$ or clumped on scales of several arc minutes to avoid detection
in the NICMOS image.  A fluctuation analysis of the new NICMOS data shows
a fluctuation spectrum consistent with that found at the same wavelength in
deep 2MASS calibration images.  The fluctuation analysis shows that the majority
of the fluctuation power comes from resolved galaxies at redshifts of 1.5
and less and that the fluctuations observed in the earlier deep 2MASS observations
can be completely accounted for with normal low redshift galaxies.  Neither
the NICMOS direct flux measurements nor the fluctuation analysis require an
additional component of near infrared flux other than the flux from normal
resolved galaxies in the redshift range between 0 and 7.  The residual 
fluctuations in the angular range between 1 and 10 arc seconds is
1-2 nW m$^{-2}$ sr$^{-1}$ which is at or above several predictions of
fluctuations from high redshift population III objects, but inconsistent
with attributing the entire NIRBE to high redshift galaxies.

\end{abstract}


\section{Introduction}

There are now two deep near infrared images of completely uncorrelated regions
of the universe.  The first is the Hubble Deep Field North (HDFN) where there 
are both very deep images at 1.1 and 1.6 $\micron$ of a 50 by 50 arc second region
\citep{thm99} and slightly shallower images at the same wavelengths of the entire
HDFN \citep{dic00}. The second is a 144\arcsec by 144\arcsec region in the Hubble 
Ultra Deep Field (HUDF) \citep{thm05} which we will call the NICMOS Ultra Deep 
Field (NUDF). The resolved object fluxes in these two fields are similar and 
10 times smaller than the 1.4 - 1.8$\micron$ Near Infrared Background Excess 
(NIRBE) found in Near Infrared Spectrometer (NIRS) on the InfraRed Telescope in 
Space (IRTS) observations \citep{mat05} and several DIRBE observations (see 
Figure~\ref{fig-nir} and \S~\ref{ss-obsi}). This discrepancy leads us to an 
investigation of the characteristics of the NIRBE and the constraints that 
the two deep fields place on it. 

Celestial observations in the near infrared, 1-2.5 $\micron$, receive flux from 
several components: stars and galaxies, zodiacal light, radiation from the telescope 
and instrument, and atmospheric emission if the observations are made from the 
ground. If there is additional flux after all of these known components are 
accounted for, the additional flux is declared an excess or in the case of the 
near infrared the NIRBE. Although the concept of a NIRBE is relatively recent, 
the literature regarding it is very extensive. For that reason we introduce the
observational background and the theoretical background in two separate subsections. 
Throughout this paper we use the convention that the flux in nW m$^{-2}$ sr$^{-1}$
is found by multiplying the flux in the band by the frequency of the center of the 
band rather than by the frequency interval of the band.

After the introduction we discuss the new NICMOS observations in the NUDF and the 
constraints they provide on the NIRBE when coupled with previous NICMOS 
observations in the HDFN. The new data more than doubles the area of very deep near 
infrared imaging in a field that is completely uncorrelated with the HDFN. 

The analysis plan is as follows. In \S~\ref{s-odr} we describe the relevant portions 
of the NICMOS HUDF observations and data reductions. We provide a detailed account 
of the background subtraction and source extraction. In \S~\ref{s-fc} we describe 
the separation of the NICMOS image flux into the individual flux components and 
compare them with the flux components found by \citet{mat05}. This is the section 
where we identify the difference between our findings and those of \citet{mat05}. 
In \S~\ref{s-flu} we turn to the fluctuation analysis of \citet{kas02} and perform 
a similar analysis on the NICMOS data which provides evidence that the fluctuations 
are due to normal galaxies with the majority of fluctuation power provided by galaxies 
in the redshift range between 0 and 1.5. In \S~\ref{s-con} we examine the constraints 
the NICMOS images place on the nature of a NIRBE. \S~\ref{s-p3} discusses the impact 
of the NICMOS images on models of the NIRBE that involve Population III stars at 
high redshift. Our conclusions are given in \S~\ref{s-conc}. Appendix A gives the 
details of the fluctuation analysis.

\subsection{Observational} \label{ss-obsi}

Information on the possibility of a NIRBE comes from a limited set of observations. 
The primary instruments are the Diffuse Infrared Background Experiment on the 
Cosmic Background Explorer satellite (DIRBE/COBE), the NIRS on IRTS, the 2 Micron 
All Sky Survey (2MASS), the Near Infrared Camera and Multi-Object Spectrometer on 
the Hubble Space Telescope (NICMOS/HST) and more recently at longer wavelengths 
with the Infrared Array Camera on SPITZER (IRAC/SPITZER). The observations up to 
2000 are covered in an excellent review by \citet{hau01} and observations since 
that time by \citet{kas05a}. For this reason we will not elaborate the previous 
observations in detail but refer the reader to these reviews. 

\begin{figure}
\plotone{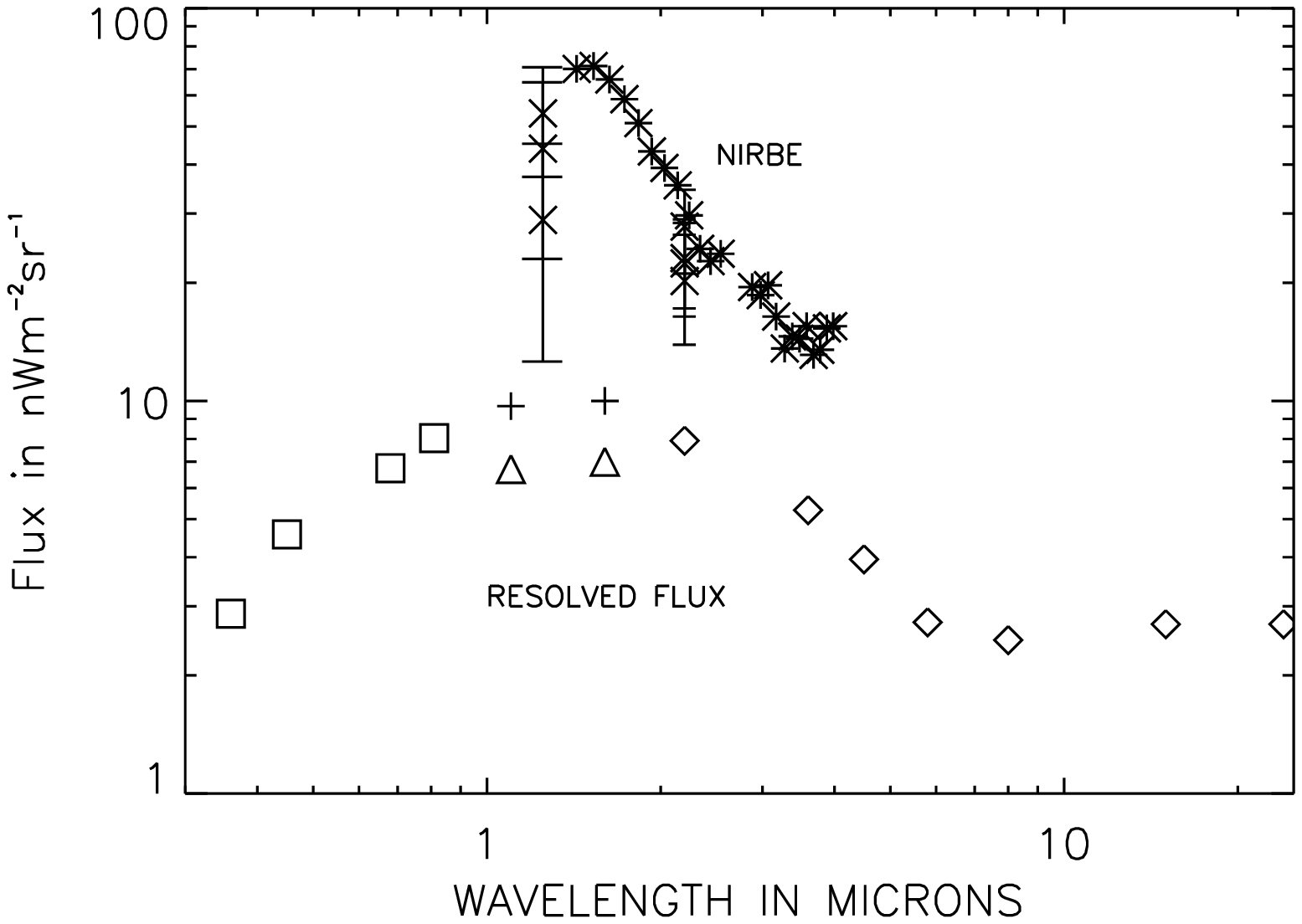}
\caption{Plot of the NIRBE flux from \citet{mat05} shown as asterisks.
The squares are object fluxes from \citet{mad00}, the triangles are the NUDF
object fluxes from \citet{thm06}, the plus signs are those fluxes corrected
for the expected missing flux from faint galaxies and regions of galaxies
and the diamonds are the quoted backgrounds from \citet{kas05a}.  The crosses
with error bars are the DIRBE fluxes from the references in table 3 of
\citet{kas05a} plus the 1.25$\micron$ point from \citet{wri01}.  Only 
claimed detections are plotted rather than upper limits. \label{fig-nir}}
\end{figure}

The primary direct evidence for a large 1.4 - 1.8$\micron$ NIRBE comes from the 
analysis of the NIRS/IRTS data by \citet{mat05}. Figure~\ref{fig-nir} shows 
their NIRBE results along with the broadband photometric fluxes at similar 
and other wavelengths from the references listed
in \citet{kas05a}. The peak of the NIRBE is at 1.4-1.6 $\micron$ and it drops to 
significantly lower levels at 3-4 $\micron$. At its peak the NIRS/IRTS NIRBE 
reaches levels of 70 nW m$^{-2}$ sr$^{-1}$. The much lower optical 
points at shorter wavelengths are from \citet{mad00} which indicate that the NIRBE 
must have a sharp cutoff at shorter wavelengths. This is the origin of the 
attribution of the NIRBE to high redshift objects. \citep{cam01} used DIRBE/COBE 
data to establish a J Band (1.25 $\micron$) point of 54.0$\pm$16.8 nW m$^{-2}$ 
sr$^{-1}$. This exceeds the upper limit of 28.9$\pm$16.3 nW m$^{-2}$ sr$^{-1}$ 
found by \citet{wri01} from the same data. These points and other DIRBE 
determinations are shown with x symbols and error bars in Figure~\ref{fig-nir}. 
An important aspect of these observations is the method of removal of the 
background and resolved object fluxes. In many cases models were used to subtract 
one or more components to find the excess flux. This is discussed further in 
\S~\ref{s-fc}. 

For the purposes of this paper we consider the NIRBE to be the excess in flux
observed by the NIRS/IRTS and DIRBE over the resolved flux shown in 
Figure~\ref{fig-nir}.  We address only the excess flux in the $1.6\micron$ region
which is primarily measured by NIRS/IRTS.

\citet{dwe05} have reviewed the direct observational evidence for a NIRBE and come 
to the conclusion that the spectrum shown by the NIRS/IRTS data is more consistent 
with zodiacal emission than with other celestial emission components. They raise 
the possibility of incomplete subtraction of the zodiacal emission accounting for 
the published NIRBE. Further evidence against the NIRBE is put forth by \citet{aha06}
who point out the observed TEV spectrum of distant blazars is too hard to have 
significantly interacted with a NIRBE at the published levels unless their intrinsic 
spectra are much harder than any known blazar spectrum. They put a limit on the NIRBE 
of $\lesssim 14\pm$4 nW m$^{-2}$ sr$^{-1}$ at 1-2 $\micron$. Similar conclusions 
were drawn by \citet{map06} with other gamma ray observations. \citet{mad00} put 
the resolved object flux at 9 nW m$^{-2}$ sr$^{-1}$ using the NICMOS HDFN observations. 

Further characteristics of the background are provided by the detection of 
fluctuations in the backgrounds observed in DIRBE/COBE images \citep{kas00}, 
2MASS images \citep{kas02}, and NIRS/IRTS \citep{mat05}. An average of the 
higher spatial resolution fluctuation analysis by \citet{kas02} of the 2MASS 
images with all detected sources removed is shown by the long dashed line in 
Figure~\ref{fig-fluct} which has a peak flux of 40 nW m$^{-2}$ sr$^{-1}$ at an 
angular scale of 1\arcsec . This fit was estimated from the 7 H band fits shown 
in Figure 1 of \citet{kas02}. The source removal is complete down to Vega H 
magnitudes of 18.7 to 19.2 which corresponds to roughly an AB H magnitude of 20. 
It should be noted that only 10 of the 4700 sources in the NICMOS NUDF image are 
brighter than an AB magnitude of 20. A more recent fluctuation analysis at 3.5 
$\micron$ by \citet{kas05b} with IRAC/SPITZER finds fluctuations peaking at 0.2 
nW m$^{-2}$ sr$^{-1}$ on angular scales of 4\arcsec - 5\arcsec. Since we do not 
have data at that wavelength we will not consider those results further other 
than to say that ruling out a NIRBE at 1.6 $\micron$ does not necessarily rule 
out one at 3.5 $\micron$.

\subsection{Theoretical}  \label{ss-thei}

Several authors (\citet{san02}, \citet{sal03}, \citet{mag03}, \citet{kas05b}, 
\citet{kas05a}, \citet{kas05c}, \citet{fer05}) interpret all or part of the 
NIRBE as the contribution from primordial Population III stars at redshifts 
between 9 and 15. \citet{san02}, \citet{sal03} and \citet{fer05} attribute
the excess flux from the NIR/IRTS observations to flux from very high redshift,
possibly population III, stars.  In this case the sharp drop off of the NIRBE 
by at least the longest wavelength WFPC2 filter in the HDFN at 8140 \AA � is 
due to the redshifted 
Lyman break and the fall off to longer wavelengths is due to the intrinsic 
spectrum of very hot population III stars. If this is true it would be a direct 
observation of emission from objects that participated in the reionization 
of the universe at a redshift of $\sim 17$ as implied by the previous WMAP results 
\citep{kog03} or at a redshift of $\sim 11$ from the recent results \citep{spe06}.
The spatial distribution of the radiation would map out the distribution of baryonic 
matter at that time and provide important constraints on the hierarchical 
clustering of ordinary and dark matter. It is important to note that under 
the interpretation that the short wavelength cutoff is due to the Lyman limit, 
the entire NIRBE flux is due to population III or at least very high redshift 
objects. In a later paper \citep{sal05} Salveterra and Ferrara raise doubts
about their earlier interpretation as is discussed below.

\citet{mag03} interpret the small angle (1-10\arcsec) fluctuations at 
$1.25\micron$ seen by \citet{kas02} as due to population III or very high redshift 
stars.  At $1.6\micron$ the contribution from population III is reduced to 
approximately 1/3 and is negligible at $2.2\micron$. Theoretical models by
\citet{coo04} and the theoretical analysis in \citet{kas04} predict fluctuations
from population III stars at levels on small angular scales that are at or below
the fluctuations observed for the all sources subtracted fluctuations presented
in Figure~\ref{fig-fluct}.  The observations presented in this publication,
therefore, do not provide a strong constraint on the validity of those calculations.  
They do bear, however, on the attribution by \citet{mag03} of 1/3 of the fluctuation 
power at $1.6\micron$ to population 3 sources.

Several theoretical objections to accounting for the entire NIRBE with 
population III stars have been raised. 
\citet{mad05} present a cogent summary of the problems. They 
point out that the energy contained in the published NIRBE requires roughly 
5$\%$ of all the baryons in the universe be converted into stars by redshift 
9 as opposed to the 2 - 3$\%$ converted into stars since that time. The efficiency 
of star formation must be on the order of 30$\%$ and the metals produced by the 
star formation must be hidden in Intermediate Mass Black Holes (IMBH) otherwise 
the metallicity of the universe would exceed solar by redshift 9. It is further 
required that accretion onto the IMBHs must be suppressed or the emission would 
exceed the observed soft x-ray background. They point out the suggestion by 
\citet{san02} that if the IMBHs could grow by accretion and form miniquasars, 
they would provide a more efficient way of producing the NIRBE. The consequences 
of this model, however, is a mass density of black holes that exceeds the 
density observed in present day galactic nuclei by 3 orders of magnitude. They 
also point out that the ionizing flux generated by population III stars to account for 
the NIRBE exceeds by a similar 3 orders of magnitude that required to produce 
the observed WMAP electron scattering depth at z = 17 quoted from the first year 
of WMAP data. 

Several authors also address the directly observable consequences of the theoretical 
models. \citet{dwe05} examine the possibility of fitting the near infrared background 
in detail. They conclude that although they can produce reasonable fits with metal 
free population III stars, a better fit is achieved with residual zodiacal light 
as discussed in \S~\ref{ss-obsi}. They do not choose between the two possibilities 
but discuss several of the reservations pointed out by \citet{mad05} to the 
population III 
model. \citet{fer05} have modeled the background with stars having a metallicity 
of 1/50 solar. They conclude that the problem of the total mass converted into 
stars is not worrisome if the majority of the mass is returned to the intergalactic 
medium and also conclude that the amount of metals returned to the interstellar 
medium can be negligible. \citet{sal05} calculate the expected flux from a cluster
containing 10$^6$ M$_{\sun}$ of population III stars at redshift 10 and find that it 
is detectable in the NUDF observations as a F110W drop out. They also calculate 
that between 1100 and 5600 such objects are required in the NUDF to account for 
the NIRBE depending on whether the collapse is by H or H$_2$ cooling. The number 
of F110W drop outs actually observed is 3 or less \citep{bou05}.

\section{Observations and Data Reduction} \label{s-odr}

A detailed description of the data reduction of the NICMOS observations in
the HUDF is given in \citet{thm05} and is not repeated here.  Only those 
items relevant to the establishment of an accurate background are discussed.  
The NUDF observations cover an area of 144$\arcsec$x144$\arcsec$ in the HUDF. 
The NUDF is located in the Chandra Deep Field South and is therefore completely 
uncorrelated with the NHDF observations.  These observations
are an independent sampling of the near infrared sky.  The NICMOS images are in
two filters, F110W and F160W, centered on wavelengths of 1.1 and 1.6 $\micron$.  The
F110W filter is very broad and is centered at a wavelength bluer than the J Band filters
used in ground based observations.  The F160W filter is almost exactly equivalent to
the ground based H filter.  The average integration time on any part of the image is
21,500 seconds.  It is deeper than the equivalent NICMOS observations in the NHDF and
is one of the deepest NICMOS images ever taken, making it an excellent image for detecting
faint sources.  The NICMOS cameras are extremely well baffled and the light entering
from angles not in the field of view is entirely negligible. 

\subsection{Background Subtraction} \label{ss-bs}

The data reduction procedure relevant to this study is the background subtraction which is 
designed to remove any instrumental background and the contribution from zodiacal light.
At the wavelengths of the NICMOS observations the primary zodiacal component is 
scattered light rather than thermal emission.  The background flux is determined by 
the median of
the individual images in each filter.  Each of the images has an area of 52 by 52
arcseconds.  The NUDF observations were taken in two epochs
separated by approximately 3 months.  The orbital position of the earth required a
rotation of the camera orientation by 90 degrees for the second epoch.  The different
sun angle meant that there were probably changes in the zodiacal emission so the median
backgrounds were compiled only from images in the same observing epoch. The images are 
spaced in a 3 by 3 grid to cover the area of the NUDF and each image position on the grid, 
which is repeated 8 times per epoch, is dithered by more than the average source size so 
that the median accurately removes the flux from resolved objects.  Inspection of the median 
images showed no trace of residual source flux.  The median images are quite smooth with
the F110W image showing a faint pattern of the flat field correction at the $1\%$ level.
This means that there is a flux component at 1$\%$ of the zodiacal flux level in the
individual images.  This is due to a slow change in the sensitivity of the camera and 
has no effect on this study.  Any study using F110W background fluctuations, however,
will have to take this into account which is why we do not use the F110W image in the
fluctuation analysis in \S~\ref{s-flu}.

The median background for each filter and epoch is then subtracted from each individual
image before the images are combined in the drizzle procedure where the original 0.2$\arcsec$
square pixels are changed to 0.09$\arcsec$ pixels to match a 3x3 binning of the 0.03$\arcsec$
pixels in the ACS HUDF images.  This procedure removes all flat background components
in the image and leaves only the contribution from resolved objects. The effect of the
median subtraction on the fluctuation analysis is discussed in \S~\ref{s-flu}.  The 
median of the sky in the final image is essentially zero as can be seen from the 
histograms of pixel values shown in Figure 3 of \citet{thm05} which has a nearly 
Gaussian distribution centered on zero.  The deviations from Gaussian noise are 
due to two reasons.  First, the drizzle procedure
produces a correlation between pixels, therefore, their distribution is not purely 
Gaussian, and second, the real sources produce a positive tail to the distribution.
The distribution underscores an important aspect of the deep fields, the vast majority of
pixels ($93\%$) sample sky, not sources.  Most of the image is zero within the noise.  This 
is what makes the median background subtraction so successful.

\subsection{Source Extraction} \label{ss-se}

The source extraction in the NUDF is described in detail in \citet{thm06} and again will
not be repeated here except for those areas relevant to this study.  Source extraction is
a two step process.  The first step is to identify all pixels that have sufficient signal
to noise to be considered part of a real source.  This is done by a process described in
\citet{sza99} which utilizes the flux in all bands.  This is done on the combined four
ACS images and the two NICMOS images.  Once the pixels have been identified a new image
is defined that has the source pixel fluxes multiplied by a large factor and all other pixels
reduced to a small random noise.  Source extraction and photometry is then accomplished
with SExtractor \citep{ber96}, hereinafter SE, in the two image mode. The first image is
the new image from the Szalay et al. procedure and the second image is the actual 
image in one of
the bands.  The first image is used to identify individual sources and the photometric 
extraction is done on the second image.  The source identification parameters in SE are
set to insure that only the pixels identified by the Szalay et al. procedure meet 
the signal to 
noise requirements.  SE then only picks sources that have the minimum number of contiguous
pixels and determines the separation in overlapping sources.  SE also returns an image
that has all pixel values equal to zero except for the identified source pixels that have
a value equal to their source ID number assigned by SE.  The total infrared power is
then calculated by simply adding up the flux in the F110W and F160W images in the pixels
that SE has identified as belonging to a source.

An important factor in this process is the relative depth of the optical images to the
near infrared images.  Due to the much longer integration time per pixel in the optical
images they go significantly deeper than the near infrared images.  In fact the majority
of identified sources would not have been identified in the near infrared images alone.  This
means that the near infrared flux is extracted in all of the areas where there are known
very faint sources and not in areas where there is no known source.  This gives a more
complete extraction of the total near infrared flux since the deep optical image guide the
extraction to the location of faint sources that would have otherwise been missed.

In addition to defining the sources, the output of the source extraction is also used to
create a source subtracted image for the fluctuation analysis described in \S~\ref{s-flu}.
In this image all source pixels are set to zero.  This image plus the weight image as
described in \citet{thm05} are used in the analysis. The total area removed by the 
source extraction if 7$\%$.

\section{Flux Components} \label{s-fc}

The primary direct evidence for a large 1.4 - 1.8$\micron$ NIRBE is the 1.4 to 
4 $\micron$ spectrum from 
NIRS on the IRTS \citep{mat05}. It is important to note that the NIRS aperture 
of 8'x12' is almost 17 times the area of the entire NUDF. Only galaxies with a 
size on the order of M32 would be resolved. In their analysis \citet{mat05} 
separated the total absolute flux into three components; the background due 
to zodiacal light and instrumental background, flux due to resolved or expected 
emission from stars and galaxies, and a remaining residual flux attributed to the 
NIRBE. In \citet{mat05} both the zodiacal and the expected emission from 
stars and galaxies were determined from models since individual objects could 
not be detected at their spatial resolution. In our image the zodiacal component 
is determined by the median of all of the images which is subtracted from all 
of the images as described in \S~\ref{ss-bs}. The detected resolved objects are 
then extracted to find the component due to stars and galaxies with estimates 
on the amount of true galaxy and star flux missed in the extraction (\S~\ref{ss-mf}). 
In our analysis there is no residual flux. To see where the analyses diverge we 
next consider all of the flux components.

\subsection{Absolute Flux}  \label{ss-af}

We first compare the absolute flux measurement before subtraction of any backgrounds
or populations to see if perhaps the NICMOS observations lie in an area of anomalously 
low intrinsic near infrared emission.  The NHDF and HUDF fields were, of course, chosen
partly on the basis of low emission due to cirrus and other sources.  However, the
observations described in \citet{mat05} and \citet{kas02} were also
taken in regions selected for low backgrounds.  The first data column of 
Table~\ref{tab-flx} 
shows the total fluxes measured in the NHDF and NUDF along with the equivalent numbers 
from \citet{mat05}. All of the fluxes listed in this table for NICMOS are fluxes from a  
flat spectrum in $f(\nu)$ that would produce the observed signal in ADUs per second.  
The fluxes for \citet{mat05} were measured from their Figure 11 and are therefore not 
exact numbers except for the residual flux which was read from their Table 1 and are 
the entries for 1.63 and 1.43 $\micron$.  The remaining
fluxes were adjusted in the last significant figure so that the sum of all of the fluxes 
equals the total flux given in the first column of fluxes.  The adjustment does not affect
the conclusions of this paper in any way. The NHDF measurements are from the observations
from Proposal ID 7817 with Mark Dickinson as PI as analyzed by \citet{thm03}. \citet{kas02} 
does not quote the total flux since that work is only interested in the fluctuation amplitude.
The NICMOS total fluxes are the detected photon rate minus the rate measured with the
cold (70 K) blank filter in place which is the ``dark''. 

The conclusion from Table~\ref{tab-flx} is that the NUDF is not an area of anomalously
low infrared emission and in fact had a higher total flux at the time of the observations than 
either the NHDF or the area of the NIRS observations.  It is also an indication of the magnitude
of the total near infrared sky intensity relative to the flux from real sources and the 
accuracy needed in subtracting out the time and spatially varying zodiacal emission.  
It is very important to note here that any differences in the conclusions on a NIRBE
do not come from differences in measured flux but in how that flux is distributed 
among the various components.

\subsection{Zodiacal Flux}  \label{ss-zf}

The second column of Table~\ref{tab-flx} is the emission due to zodiacal light.  As described
in \S~\ref{ss-bs} the calculation for the NICMOS images assumes that all contributions to
the median image are from the zodiacal light and any instrumental background.  The zodiacal
emission in \citet{mat05} is calculated from the models of \citet{kel98}.  The zodiacal
levels measured in the NICMOS observations are about 100 nW m$^{-2}$ sr$^{-1}$ higher
than the calculated values used in \citet{mat05}.  This is the primary variance in the two
analyses and is the probable origin of the claim of a NIRBE by \citet{mat05}.  This is
particularly relevant to the finding by \citet{dwe05} that the spectrum of the NIRBE flux 
published by \citet{mat05} is better fit by zodiacal emission than by flux from high redshift
population III stars.  Note that the higher 1.6 $\micron$ zodiacal flux quoted for the NUDF
probably includes a component of instrumental background.  The NHDF observations were
taken during the cryogenic operation of NICMOS while the NUDF observations were taken
during the warmer operation with the NICMOS Cooling System (NCS).  The F160W filter
extends to 1.8 $\micron$ which is subject to thermal emission from the warm instrument
optics.

\subsection{Flux due to Observed Sources} \label{ss-os}

The NUDF and NHDF resolved object flux at 1.6 and 1.1 $\micron$ is listed in the detected or
expected sources column of Table~\ref{tab-flx}.  The detected object flux for both fields
and both filters is about 7 nW m$^{-2}$ sr$^{-1}$ which is a factor of 10 less than the
published NIRBE.  It is consistent with the value of 9 nW m$^{-2}$ sr$^{-1}$ for the NHDF 
found by \citet{mad00} from the same NHDF data but with a different data reduction.
The NHDF and NUDF resolved object fluxes are a factor of 5 lower than the stellar flux 
listed for NIRS.  The NIRS galaxy and star flux is calculated from from the ``SKY'' model
of \citet{coh97} since NIRS does not resolve most objects.  The lower object flux in the
deep fields may not be inconsistent with the model since the deep fields were chosen to avoid 
bright stars and galaxies.  The number of objects detected in the NUDF is about 4700, of 
which about 1/3 have flux above the detection threshold in the F110W and F160W NICMOS bands.
The positive error of 3.0 nW m$^{-2}$ sr$^{-1}$ is due to the possible missed flux from
faint sources discussed in \S~\ref{ss-mf}.  The 0.3 negative error indicates that the
source extraction is relatively conservative and that the field was rigorously checked
for spurious sources which were then rejected.  The detected fluxes come from over 600,000
pixels so the Poisson error is quite low.  The primary source of this error is the absolute 
calibration of the NICMOS sensitivity which is accurate to $5\%$ or better.

\subsubsection{Flux from Undetected and Faint Outer Parts of Galaxies} \label{ss-mf}

For completeness we can ask how much flux has been missed and resides in pixels that have 
less flux than the detectable limit.  This is a legitimate question since the the number
of pixels below the cutoff limit is 20 times the number of source pixels.  Figure~\ref{fig-fp}a
shows a histogram of the number of source pixels having a given flux in nJy.  
Remarkably the slope
of the histogram in the log-log plot is almost exactly $-1$.  This translates to a log normal
flux distribution where the flux per dex is constant.  This is log divergent on each end
of the distribution.  The bright end is, of course, cutoff as the distribution abruptly becomes
much steeper.  The faint end is terminated when the number of pixels in the distribution 
equals the number of pixels in the image.  If the $-1$ slope is extended to $1.3 \times 10^{-2}$
nJy all of the pixels will be accounted for.  The faint end distribution deviates from 
the $-1$ slope at 1 nJy.  Integration of the distribution between $1.3 \times 10^{-2}$ and
1 nJy yields less  than 1/2 of the flux between 1 and $10^4$ nJy. This can only 
increase the actual flux by $50\%$ even if the actual detections at less than 1 nJy are 
ignored.  This analysis assumes that the true faint object and pixel slopes do not deviate
from the slopes determined in Figure~\ref{fig-fp}.

We can also do the same calculation for detected sources as shown in Figure~\ref{fig-fp}b.
The slope in this figure is $-0.63$ which gives a lower correction than the power per pixel.
The shallower slope make physical sense since the fainter parts of galaxies cover more area
and hence more pixels than the bright regions.  As a result of these distributions we put
the amount of missed flux equal to or less than $50\%$.  This is the origin of the two NICMOS
points in Figure~\ref{fig-nir} which are marked as crosses above the triangles which 
represent the resolved object flux actually detected.  The correction for missed flux is
consistent with the corrections to the star formation rate calculated in \citet{thm03} and
\citet{thm06} by different means. It should also be noted that confusion is not a
factor since only $7\%$ of the pixels contain source flux.

\subsection{Residual Flux or NIRBE}

The assignment of fluxes between zodiacal and resolved objects accounts for all 
of the observed flux in the NICMOS observations in the deep fields.  The flux 
not in detected objects or in the median background is $0.0^{+3.}_{-0.3}$ nW 
m$^{-2}$ sr$^{-1}$ from \S~\ref{ss-mf} above. In the NIRS observations discussed
in \citet{mat05} the modeled zodiacal and object fluxes come up short of the total flux.
The remaining flux is declared a residual flux which is the NIRBE.  Our analysis
determines that the most likely explanation is that the models did not have the required 
accuracy to account for the flux components and that the published NIRBE is most likely
residual zodiacal flux that was unaccounted for by the zodiacal model. The same 
arguments can also be made for the DIRBE observations that generally lie below the
NIRS fluxes.  Although in some cases \citep{wri01} galaxy and stellar sources have 
been removed by referring to 2MASS images, the zodiacal flux is still determined
by a model.  Again it should be noted that there is not a discrepancy in the total
observed flux, but rather in the way it is distributed between the various emission
components. There are at least
two ways, however, where the NICMOS analysis would not discover a NIRBE.  The first is if
the NIRBE is very flat and mistaken for part of the zodiacal flux. The second is if
the NIRBE is very clumped and missed by the small deep fields.  These possibilities 
are discussed further in \S~\ref{s-con}.

\section{Fluctuation Analysis} \label{s-flu}

We next address the origin of the fluctuations found in the 2MASS deep calibration 
fields by \citet{kas02}. The 2MASS fluctuation spectrum extends from scales of 
1$\arcsec$ to 100$\arcsec$.  As mentioned in the introduction an eyeball estimate
of the average spectrum is given by the dashed line in Figure~\ref{fig-fluct}.  
The limiting magnitude for source removal from the 2MASS images is approximately 
H = 19 in Vega magnitudes which is roughly an AB magnitude of 20.  There are only 10
sources in the NUDF brighter than an AB magnitude of 20.  The detected limiting AB
magnitude in the NUDF is approximately 28.5.  This provides the opportunity to 
check whether the fluctuation spectrum can be accounted for by the observed sources
which extend to a maximum redshift of 7 or whether the fluctuations require a new 
population of high redshift sources.

We performed a fluctuation analysis on 5 images, i) the NUDF F160W image, 
ii) the image with the 10 sources brighter than 20 mag AB removed to simulate the 
source subtracted 2MASS images, iii) the image with all detected sources removed, 
iv) an image created by drizzeling the first and second epoch median images in 
exactly the same way as the true images, and v) an image of gaussian noise at the
level of the NUDF image noise. The fluctuation analysis procedure is 
described in detail in Appendix A. The only operation performed on the images 
before the Fourier transform is to set any small DC 
level to zero to prevent ringing due to a DC component. Figure~\ref{fig-ims} shows 
a small portion the NUDF image with all of the sources removed. The results of the 
analysis are presented in Figure~\ref{fig-fluct}. The error bars shown in the
figure are for a Gaussian noise due to large scale structure calculated as 

\begin{equation}
\Delta P_2 = \frac{P_2}{\sqrt{N_k}}
\label{eq-gn}
\end{equation}

\noindent where $N_k$ is the number of k components in the half ring of $\Delta k$
used to define the wavenumber bins in Figure~\ref{fig-fluct}. We stress that although 
Gaussian noise may be appropriate for large scale structure, it is an underestimate 
of the effects of shot noise. The fluctuation power in the Figure has been adjusted
for the small amount of sky blanked by removing the sources.

At the request of the referee we computed the fluctuations in the image with the 
10 sources brighter than 20 mag AB removed ignoring the regions along the axes
in Fig.~\ref{fig-2d} that correspond to angles of 26 arcseconds or greater.  This
removes some of the points at larger wavenumbers as well.  The only
effect, other than to remove the wavenumbers corresponding to 26 arcseconds or greater,
was to lower the point in Fig.~\ref{fig-fluct} corresponding to approximately 18
arcsecond by an amount equal to the height of the diamond symbol. 

The median image power spectrum gives an indication 
of how much the fluctuation spectrum can be affected by power in the median image.  
Figure~\ref{fig-fluct} shows that at angular scales of 10 arcseconds or greater there 
can be an influence on the all sources subtracted fluctuation spectrum from the 
median subtraction.  The majority of power from resolved galaxies, however, is at 
smaller angular scales and any median subtraction effects do not alter the 
conclusions of this paper.  It should be emphasized that our conclusion that we would 
miss backgrounds that are flat on scales of 100 arcseconds comes from our 
analysis in \S~\ref{ss-dif}, not from the fluctuation spectrum.  We also emphasize
again that we are not sensitive to fluctuations from population III stars at the
pessimistic levels calculated by \cite{coo04} and only at the levels calculated
by \citet{kas04} for objects formed at redshift 10.

The amplitude of the two dimensional power spectrum, $\sqrt{f(u,v)f^*(u,v)}$ in 
the notation of the appendix, of the source subtracted image in Figure~\ref{fig-ims} 
is shown in Figure~\ref{fig-2d}. All quadrants are shown although the power spectrum is 
symmetric about the horizontal axis.  The stretch is given in the figure caption.
Figure 5 shows that there are no significant artifacts in the NUDF image. The 
small dots and regions of higher intensity are most likely due to the drizzle 
procedure used to convert the original NICMOS camera 3 0.2 arc second pixels to 
the 0.09 arc second pixels used to match the rebinned ACS pixels as described in 
\citet{thm06}. Due to the small area removed by the source subtraction, no 
correction of the power spectrum for the lost area was made.  

The difference between the fluctuation spectrum of the full image and the 
source subtracted image shows that the NUDF contains a rich spectrum of 
fluctuations due to the sources in the field.  When sources brighter than 
20 mag (AB) are removed the fluctuations are reduced, with a similar 
amplitude and shape to that of \citet{kas02}. When we further remove all of 
the sources detected in the NUDF the fluctuation is reduced to a level similar 
to that of a Gaussian noise field. This clearly shows that the fluctuations 
from detected NUDF sources with redshifts between 0 and 7 are fully capable of 
accounting for the fluctuations observed in the 2MASS images without invoking 
a new population of sources.  The vast majority, if not all, fluctuations are 
simply due to galaxies fainter than the 2MASS calibration field limit which are, 
however, easily detected in the NUDF image. It is also relevant to the presence of 
a NIRBE regardless of the type of objects invoked. It shows that \emph{normal sources 
with a total flux of 7 nW m$^{-2}$ sr$^{-1}$ reproduce the fluctuations observed 
in the deep 2MASS calibration images}. Sources with a total flux 10 times that 
amount, distributed similarly as the observed sources, would overproduce the 
fluctuations.

As shown in the work of \citet{coo04} the present fluctuation analysis is not 
sensitive enough to measure the fluctuations expected from a pessimistic
assumption on the expected fluctuations from population III sources.  The difference
between the pessimistic and optimistic assumptions is 5 orders of magnitude
in power and 2.5 orders of magnitude in fluctuations.
A comparison with the predictions in \citet{kas04} is presented in \S~\ref{s-p3}. 
It is inconsistent, however, with high redshift sources of any kind that provide 
the same flux as ascribed to the NIRBE fluxes claimed in \citet{mat05} as was
investigated by \cite{mag03}. In that work 1/3 of the H band fluctuation power at small 
angles is due to population III sources which is clearly not the case in 
Figure~\ref{fig-fluct} where the all source subtracted fluctuations at an angular
scale of 5 arc seconds are factor of 30 below the magnitude 20 source subtracted fluctuations.
If the sharp drop in flux between 1.4 $\micron$ and optical flux is interpreted as 
due to the Lyman limit of high redshift sources then all of the observed flux must 
be due to such sources. The fluctuation analysis presented here is then evidence 
against that interpretation.

The NICMOS fluctuations diverge from the noise spectrum at an angular scale of 
10\arcsec. It is not clear whether the NICMOS fluctuations at these scales are 
due to faint sources or to other factors such as power in the subtracted median
image. It should be noted that the fluctuations are on a scale consistent with 
small flat fielding errors. We have not attempted to analyze the source subtracted 
NUDF fluctuations in detail by determining the components due to noise, incomplete 
flat fielding, median subtraction and faint galaxies. The source 
subtracted fluctuation amplitudes are therefore only upper limits on fluctuations 
due to light not coming from the resolved sources. It is also evident from the 
difference in the fluctuation amplitude for the complete image and the image with
the 10 brightest sources removed from the 4700 total sources, that the fluctuation
amplitude can vary markedly from field to field, depending on the number of 
``bright'' sources included.

\subsection{Fluctuations versus redshift} \label{ss-fz}

We next look at contributions to the fluctuations as a function of redshift. To 
do this we created a series of images which had all sources removed except those 
in a given redshift range and repeated the analysis. The first redshift range is 
z =0.0-0.5, with subsequent ranges of width 1 centered on integer redshifts from 
1 to 6 and a final range of redshifts above 6.5. The redshifts are photometric 
redshifts taken from \citet{thm06}. As such they are subject to uncertainties 
but photometric redshifts have been shown to be reliable with occasional catastrophic 
failures. Most of the redshift ranges contain several hundred galaxies so occasional 
catastrophic failures should not alter the bulk conclusions. The number of galaxies 
in each redshift bin is given in Table~\ref{tab-nz} along with the flux from the 
galaxies in the bin. Note that although the last bin extends 
to redshift 10, the limit of the photometric redshift procedure, the detected 
galaxies are in the redshift range between 6.5 and 7. The fluctuation plots are 
presented in Figure 6. It is clear that most of the fluctuation power comes from 
galaxies in the redshift range between 0 and 1.5. By a redshift of 4 the 
fluctuation power is essentially the same as the all-source-removed fluctuations 
indicating that the resolved sources at redshifts of 4 and beyond contribute very 
little to the observed fluctuations.

\section{Constraints on Possible Cosmic Near Infrared Backgrounds} \label{s-con}

The combined NUDF and NHDF fields cover a little over 11 square minutes of sky, in two
almost equal areas, in two completely uncorrelated regions of the sky.  They contain
several thousand objects in areas chosen for a low density of objects.  The resolved
objects, corrected for missed faint flux, have a maximum total flux of 10 nW m$^{-2}$ 
sr$^{-1}$ at 1.1 and 1.6 $\micron$.  This is 7 times less than the NIRBE 
measured by \citet{mat05}.  The NICMOS Hubble deep field observations clearly demonstrate 
that \emph{if a NIRBE exists it can not come from luminous objects that have a spatial
distribution of mass and light similar to baryons at redshifts of 6 and less}.  
There is no way to hide 7-10 times the observed flux in the two fields in spatially 
resolved objects.  Any source of a unresolved background over the observed power in 
galaxies or stars must either be flat or clustered.  A flat 
source of background, not coming from the zodiacal light, will be removed by our median 
subtraction.  If the source of a background is clustered, the two deep fields 
may have missed observing them. The objects causing the NIRBE would have to be clustered
in a manner that the average space between them is more than 2 to 3 arc minutes, but
in a way that, while emitting 7-10 times the power of the objects observed in the
two deep fields, they avoid detection as resolved objects in the wide field of view
DIRBE, NIRS and 2MASS observations.  

Our fluctuation analysis of the source subtracted image does not directly rule out an 
isotropic background.  However, within the concordance $\Lambda$CDM model, high
redshift galaxies will be clustered, and their emission will show angular
anisotropies.  Limber's equation \citep{lim53} allows one to predict
the angular clustering of sources from the spatial power spectrum.
The linear power spectrum with a scale independent bias is a conservative lower limit on the
fluctuations, and at high redshift it is not a bad estimate.  With the 3-d power spectrum, 
the Limber projection is standard \citep{bau94}.  Neglecting curved sky effects (which 
are very negligible here) and assuming a flat cosmology, one has

\begin{equation} 
P_2(K) = I_{NIRB} \int {dr\over r^2} P_3(K/r) f^2(r)
\label{eq-p2k}
\end{equation}

\noindent where $K$ is the wavenumber (aka, $\ell$), $r$ is the comoving coordinate 
distance (i.e. $\int (c/H)dz$), $I_{NIRB}$ is the average celestial Near Infrared 
Background of 7 nW m$^{-2}$ sr$^{-1}$ and $f(r)$ is the distribution of the
sources as a function of $r$, normalized to unit integral. $P_3(k)$ is the spatial
power spectrum in comoving coordinates, normalized as described in the following
paragraph.  At high redshift,
the slices are reasonably thin in $r$ compared to the distance out to
that redshift, so one can approximate $r$ as slowly varying over the
integral of $f^2$.  For a uniform distribution of sources between two
redshifts, the integral of $f^2$ is simply $1/\Delta r$, where $\Delta r$
is the thickness of the slab in $r$.  More peaked distributions would
give larger answers, increasing the angular correlations.  Hence, our
lower bound on $P_2(K)/I^2_{NIRB}$ is $P_3(K/r)/r^2\Delta r$.  

We parameterize the amplitude of the linear power spectrum by $\sigma_{8,flux}$,
which is the rms fluctuation of the emissivity in 8h$^{-1}$ comoving Mpc
radius spheres for the redshift in question.  This value is also the $\sigma_8$
for galaxies if one weights the galaxies by their luminosity.
With this, we find that the rms fluctuations, $\mathcal{F}_K$ in 
Appendix A, at $2\pi/K=10''$ scale are $0.1 I_{NIRB}\sigma_{8,flux}$ for a screen of 
sources between redshifts 10 and 14. In other words, for $\sigma_{flux,8}=1$, 
we predict that the power at $10\arcsec$ scale is $10\%$ of the angle averaged
level of the background, if that background is generated at $10<z<14$.
The fluctuations scale as the square root of the thickness of the slab;
increasing the range to $6<z<20$ only halves the rms fluctuations. Shot
noise or non-linear structure formation would increase the clustering above 
the lower limit set by the linear power spectrum.  While we are quoting our 
amplitude by the familiar $\sigma_8$, the scales probed by these data are closer 
to 0.3Mpc and the linear matter power spectrum is used to make the extrapolation.
Small scale filtering by the IGM could cause the smal scale fluctuations to be 
less than the linear prediction. 

The ratio of our observed $10\arcsec$-scale fluctuations to the DC NIRBE level
of \citet{mat05} is only about $0.7\%$ rather than the $10\%$ found above.  To 
be consistent with the entire NIRBE being due to high-redshift sources the
sources would have to be 
surprisingly unclustered ($\sigma_{8,flux}\lesssim0.1$ even with shot noise) or 
the emission must be scattered to smooth out fluctuations on $10\arcsec$ scales, 
e.g., by scattering of Lyman $\alpha$ into halos much larger than $10\arcsec$.  
The emission-weighted bias of the undiscovered high redshift galaxies is of 
course not known, but known populations of high-redshift galaxies are highly 
clustered (\citet{ade05}, \citet{coi04}) and the rare density peaks that form 
luminous very high redshift proto-galaxies are expected to be highly biased.
 
Generating the background in thicker
slabs at lower redshift can produce smaller fractional fluctuations,
as there is more dilution along the line of sight.  However, it is
implausible that there would be 10 times more emission at low redshift
than the sources we already detect, given the extreme depth of the NUDF
data and the budget of normal stars in the local Universe. In other words, 
barring significant scattering, the low level of our observed fluctuations 
place an upper limit of roughly $10/\sigma_{8,flux}$ nW m$^{-2}$ sr$^{-1}$ 
on the flux from high redshift objects (for the $6<z<20$ assumption).  Of course,
the value can be lower if the observed level is due to noise (as is probably
the case), instrumental effects, or the clustering of incompletely removed 
low-redshift sources.

\section{Constraints on a Population III NIRBE} \label{s-p3}

As discussed in the introduction, redshifted light from population III stars 
has been the topic of several theoretical investigations into the possible origin of a
NIRBE.  At the proposed redshifts these objects will appear in the NICMOS F160W band
but not the F110W and be labeled as F110W dropouts.  The total number of F110W dropouts
in the NUDF is 3 or less \citep{bou05}. Note that a detection in a single band can not
distinguish high redshift objects from extremely red objects.  We will divide
our constraints on population III into fluctuation constraints and direct flux
constraints. 

\subsection{Fluctuation Constraints}

Predictions of the fluctuations from population III objects have been made by
\citet{mag03}, \citet{kas04} and \citet{coo04}.  \citet{mag03} ascribe all of 
the fluctuations observed in the $1.25\micron$ 2MASS deep calibration images 
after source subtraction to population III objects which is reduced to 1/3 at
$1.6\micron$.  As mentioned previously our fluctuation analysis of the source
subtracted NUDF images produces an upper limit on fluctuations at the 5 arc
second scale that is 1/30 of the 2MASS fluctuations, inconsistent with the
\citet{mag03} prediction. We next consider the predictions of \citet{kas04}
using the z formation of 10 models from Figure 5 of the paper.  In this work
only 45 nw m$^{-2}$ sr$^{-1}$ are ascribed to CIB flux at $1.25\micron$ and
the fluctuations at a scale of 5 arc seconds are about 5 nw m$^{-2}$ sr$^{-1}$.
Following \citet{mag03} we assume that the fluctuations at $1.6\micron$ will
be 1/3 that or 1 to 2 nw m$^{-2}$ sr$^{-1}$.  The observed NUDF fluctuations
at 5 arc seconds are 1 nw m$^{-2}$ sr$^{-1}$, consistent with the prediction,
particularly given possible variation in fluctuation power in different small
fields.  The power predictions from \citet{coo04} vary over 5 orders of 
magnitude in power and 2.5 orders of magnitude in fluctuations
depending on the assumptions.  The vast majority of that space is far below
the sensitivity limit of these observations.  In all of the predictions the
primary population III fluctuation power is at larger angular scales where 
the galaxy fluctuation power should be significantly less.  The bottom line
is that any contribution from population III stars must have a fluctuation
power at small angular scales that is equal to or less than the fluctuation
power of the all sources subtracted angular distribution shown in 
Figure~\ref{fig-fluct}.

\subsection{Direct Flux Constraints}

In the following we consider two types of population III emission; point or resolved 
emission and diffuse emission from scattered Ly $\alpha$ photons.

\subsubsection{Point or Resolved population III Sources} 

Our fluctuation analysis clearly shows that you can not distribute 10 times the flux
of the detected objects in additional objects similar to the detected galaxies. At
$1.6\micron$ \citet{mag03} calculate that 1/3 of the observed 2MASS fluctuations 
are contributed by population III objects. At an angular scale of 5 arc seconds
the all source subtracted fluctuations are a factor of 30 below the 2MASS fluctuations 
and a factor of 100 below the observed NUDF fluctuations from all of the sources.
We will, nevertheless, consider the detectability of individual population III galaxies 
as discussed in the literature. As mentioned in the introduction, \citet{sal05} 
calculate the flux from a $10^6$ M$_{\sun}$ cluster of population III stars
at a redshift of 10 and find a spectral peak at 1.6 $\micron$ of 60 nJy with
an average of 40 nJy in the F160W band.  Figure~\ref{fig-fp} indicates that the 
NUDF is complete up to sources of 25 nJy so any such source is easily detectable
in the NUDF F160W image.  They further calculate that to produce the published NIRBE
each NUDF sized region would contain 1165 sources (under the assumption of H cooling) 
to 5634 sources (under the assumption of H$_2$ cooling). This far exceeds the 
observed number of 3 or fewer.  Even if the
sources are more realistically arranged in a distribution of masses there should be
far more detections than observed unless the upper mass limit on the population III
stars in the clusters is a few times $10^5$ M$_{\sun}$.  This calculation, 
however, assumes that the emitted light is distributed in the same manner as 
the mass.

\subsubsection{Flux from Diffuse Ly$\alpha$ Emission} \label{ss-dif}

Even if the proposed Pop III stars are spatially distributed in a manner similar to
lower redshift stars, their light may be differently distributed.  Most of the photons
from a massive, hot Pop III star are emitted beyond the Lyman limit and converted
into Ly$\alpha$ photons that are extensively scattered by the surrounding neutral
gas \citep{loe99}.  Most of the background power is then emitted in a diffuse area 
surrounding the star.  Galaxies made up of these stars would have apparent angular 
sizes between 10 and 100 arc seconds \citep{kob06}.  To test our sensitivity to a 
diffuse emission, still spatially distributed similar to baryons at lower redshift, 
we created a simulated image of diffuse emission.  The initial image placed 
diffuse emission in a Gaussian shape with FWHM = $10\arcsec$ at the location of 
every detected object in the NUDF image. The photon flux of each object was 
multiplied by 10 to account for the published NIRBE. The simulated image was 
reflected about both the x and y axis to produce a ``random'' object field.  
The diffuse NIRBE image was then added to the original field to see if the 
simulated NIRBE was detectable.  The simulated NIRBE was easily detectable as 
diffuse but resolved objects, particularly due to the small number of NIRBE 
sources from bright objects that stood out as intense diffuse objects.

The probability of any high redshift object being 10 times as bright as the 
brightest object in our field is exceedingly remote.  The most efficient way 
to hide the flux from detection is to distribute it equally among all of the 
objects.  A second image was created with all of the objects having equal flux 
with a total flux equal to the published NIRBE.  The median was subtracted from 
this field as would happen in our reduction and the simulated NIRBE again added 
to the observed image.  The results are shown in Figure~\ref{fig-bac} where the 
images with and without the NIRBE are compared.  Again the NIRBE sources are easily
detectable as enhanced areas of emission.  Finally the procedure was repeated using
Gaussian sources with a FWHM = $100\arcsec$ which is nearly the size of the image.  
As would be expected this case produced a very uniform NIRBE which is subtracted 
out during the median subtraction routine and would have been attributed to 
zodiacal light in our image processing procedures. If, however, the constraint 
of exactly equal flux for all objects is lifted, the NIRBE flux is again detectable.
These tests indicate that the only way a NIRBE at the flux levels indicated by
\citet{mat05} could be not detected by our observations is if it is flat on the
scale of $100\arcsec$ or larger or if it is clumped on the scale of several arc
minutes and our two fields missed the objects contributing to the NIRBE.

\section{Conclusions} \label{s-conc}

A fluctuation analysis of the NUDF clearly shows that the fluctuation power found
in the 2MASS fields by \citet{kas02} is easily provided by sources with redshifts
less than 7 that are below the 2MASS detection limit but easily detectable in
the much higher signal to noise NUDF image.  There is no need to invoke a high
density of population III stars at high redshift to account for the observations. In 
fact most of the power in the fluctuations comes from low redshift rather than high 
redshift sources. More important it also shows that sources with a total flux 7 to 
10 times less than the flux attributed to the NIRBE produce fluctuations equal 
to the fluctuations observed in the deep 2MASS calibration images. Sources with 
fluxes equal to that attributed to the NIRBE, spatially distributed in the same 
way as the resolved galaxies would overproduce fluctuations by a factor of 7 to 
10. The residual fluctuations after source subtraction put a reasonable upper
limit of $10\sigma_{flux,8}$ nW m$^{-2}$ sr$^{-1}$ on any unresolved background 
component spatially
distributed in a manor similar to the observed baryons in the universe. It may be
possible to flatten the emission component if it is essentially all Lyman $\alpha$
emission that is scattered over $100\arcsec$ or more but the analysis presented here
indicates that there is no observational or theoretical evidence that requires such 
a component.

In regard to the NIRBE observed by \citet{mat05}, two of the deepest images ever 
observed at 1.1 and 1.6 $\micron$ fail by a factor
of 7-10 to find resolved objects capable of providing the flux levels of the published 
1.4 - 1.8$\micron$ NIRBE.  Although the absolute power observed by NICMOS and NIRS are
consistent, the distribution of that power between a flat component and resolved sources 
differs.  The very high resolution NICMOS images differentiate the two components by direct 
observation, unambiguously separating the resolved component from the flat component.  
The separation in the NIRS observations is accomplished with models for both the zodiacal
light and the stellar and galaxy population. The amount of flux attributed to the 
zodiacal light by the model used in the NIRS observations is significantly less than
the flat background found in the NICMOS observations and the difference is of the 
same order of magnitude as the NIRBE. This leads to the possibility that the flux 
attributed to the NIRBE by \citet{mat05} is simply residual zodiacal light 
unaccounted for by the model.  This is reinforced by the observation by \citet{dwe05} 
that the spectrum of the NIRBE found by \citet{mat05} is very similar to the spectrum 
of zodiacal light, as was also noticed by \citet{mat05}. Since the high z population III 
light should have a similar spectrum this is not definitive evidence for a zodiacal 
interpretation but it does indicate that the data is consistent with a zodiacal 
interpretation.

These observations, coupled with the theoretical implications of the published
NIRBE discussed in \S~\ref{ss-thei} lead us to the conclusion that the NIRBE
does not exist and arose from an incomplete subtraction of the zodiacal light.  An
important caveat to this conclusion is that the two deep NICMOS field cover an extremely
small area of sky.  This means that any NIRBE that is flat such as might be
produced by widely scattered Ly $\alpha$ emission, or is clumped, as perhaps
population III stars might be, will either be subtracted from or missed by the NICMOS deep
fields. These caveats, however, do not dismiss the theoretical objections to the NIRBE.

\acknowledgments
This article is based on data from observations with the 
NASA/ESA Hubble Space Telescope, obtained at the Space Telescope Science 
Institute, which is operated by the Association of Universities for Research 
in Astronomy under NASA contract NAS 5-26555.  RIT, DE, and XF are 
funded in part by NASA Grant HST-GO-09803.01-A-G from the Space Telescope 
Science Institute.

\appendix \label{a-fluct}

\section{Calculation of Fluctuations in the NUDF} \label{s-fluc}

In order to compare our observations with the fluctuations observed in the 2MASS
images by \citet{kas02} we performed a fluctuation analysis on the NUDF
F160W image.  Due to the small scale flat field residuals described in 
\S~\ref{ss-bs} we did not perform a similar analysis on the F110W image.
The purpose of this appendix is to give an exact description of the analysis
method so that it can be repeated on the NUDF Treasury images in the HST
archive.  For ease in performing the Fourier transforms, the mosaic image
in its camera oriented rectangular format was used rather than the rotated
version with north up.

\subsection{Source Removal}

All pixels identified by SE as part of a source are set to 0 for the source
subtracted image and only those from sources brighter that 20 AB mag for the
comparison image to the previously analyzed 2MASS images.  The procedure
for identifying those pixels is described in \S~\ref{ss-se}.  Next the weight
image from the Version 2 NUDF Treasury submission was divided by its median
to produce a normalized weight image.  The source subtracted image was then
multiplied by the normalized weight image.  After this step the median of the 
image was subtracted from the image to remove any remaining residual
DC component before performing a 2 dimensional Fourier Transform.  No further
processing was done on the image.

\subsection{Fluctuation Analysis}

The fluctuation analysis follows the prescription for analysis of fluctuations
in the temperature of the CMB presented in \citet{pea99}.  The original image
which is in ADUs per second is converted to Watts m$^{-2}$ Hz$^{-1}$ per pixel
by the standard conversions for the NICMOS instrument.  The flux is then multiplied
by the frequency of 1.6 $\micron$ and divided by the solid angle in steradians
subtended by a single pixel to produce a new image in units of Watts m$^{-2}$ 
Sr$^{-1}$ which is designated $F(\vec{\theta})$ where $\vec{\theta}$ is the angular distance
from the lower left corner of the image in radians.  We then take the Fourier 
transform of the image defined as 

\begin{equation} 
f_K(\vec{K}) = \frac{1}{L^2} \int F(\theta) e^{-\vec{K} \cdot \vec{\theta}}d^2\theta
\label{eq-ft}
\end{equation}

\noindent  where $L$ is the angular dimension of the square image and $\vec{K}$ is
the wave number.  The square of the fluctuation for wave number $\vec{K}$ is given
by

\begin{equation}
\mathcal{F}^2_K = \frac{L^2}{(2\pi)^2} 2\pi K^2 \mid f_K \mid^2
\label{eq-fl}
\end{equation}

If $\vec{K}$ is in units of inverse radians, then it is equivalent to the multipole 
degree $l$ and

\begin{equation}
\mathcal{F}^2_K = \frac{l^2}{2\pi}C_l
\label{eq-cl}
\end{equation}

\subsection{Translation into Fourier Series}

Although equations \ref{eq-ft} and \ref{eq-fl} provide elegant definitions for
the fluctuation, the actual calculations involve digital Fourier series rather
than integrals.  We used the IDL Fourier series procedure in this analysis so 
we will use that definition in this appendix.  The IDL forward Fourier series is
defined by

\begin{equation}
f(u,v) = \frac{1}{N^2} \sum_{x=0}^{x=N-1} \sum_{y=0}^{y=N-1} F(x,y) e^\frac{-2\pi iux}{N} e^\frac{-2\pi ivy}{N}
\label{eq-ift}
\end{equation}

\noindent The returned wave numbers are given by

\begin{equation}
u_i = \frac{i}{N\delta \theta}
\label{eq-iwv}
\end{equation}

\noindent where $\delta \theta$ is the pixel spacing in radians.  Comparing 
equation~\ref{eq-ft} with equation~\ref{eq-ift} we note that $L=N \delta \theta$
and that $d^2\theta = \delta\theta^2$ giving the $\frac{1}{N^2}$ term in the IDL
Fourier Series.  We also note that the components of $\vec{K}$ in the x and y
directions are $2\pi u$ and $2\pi v$.

\begin{figure}
\plotone{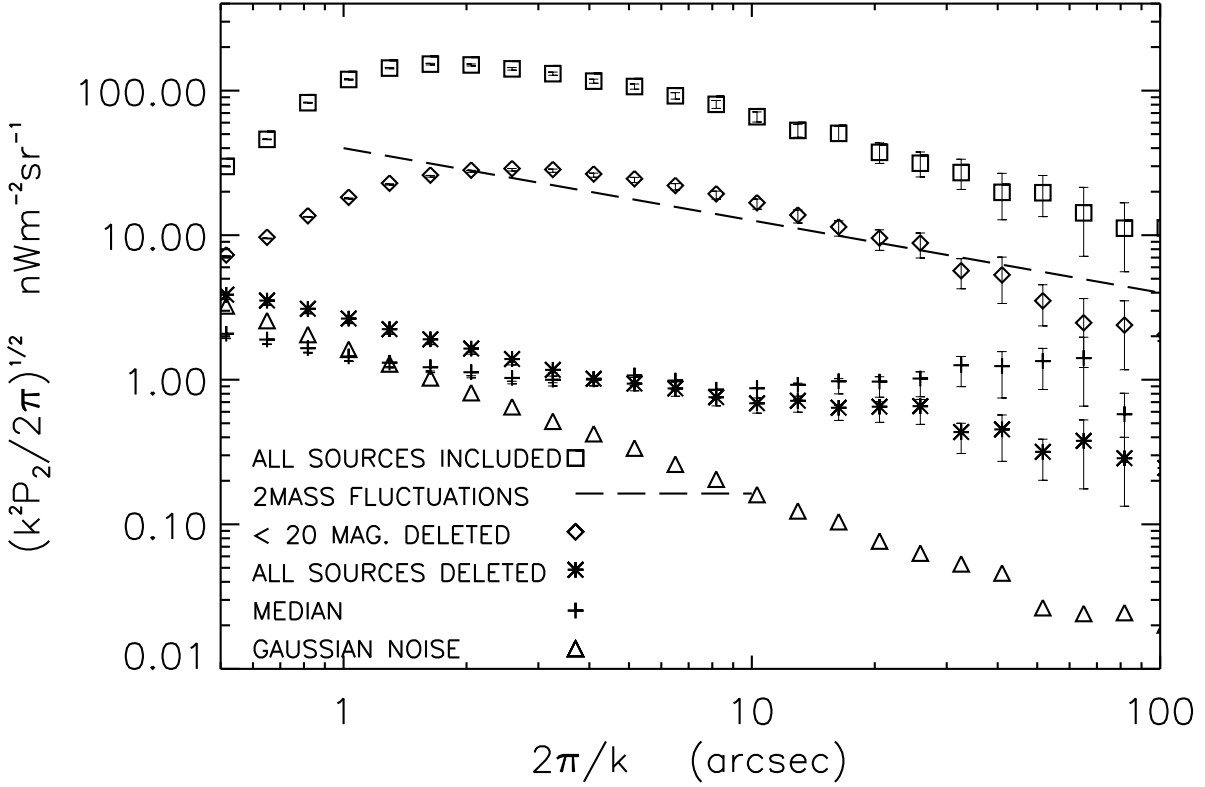}
\caption{The fluctuation spectrum of the of the F160W NUDF image is given
by the squares, the image with sources brighter than 20 AB mag. subtracted by
the diamonds, with all sources subtracted by the asterisks, and the fluctuations
of a Gaussian noise field by the plus signs. The dashed line represents an
average of the fluctuations found by \citet{kas02} in 7 different 2MASS
calibration fields.  The photon Poisson noise for the all sources included
and brighter than 20. mag deleted curves is smaller than the symbol sizes.
The error bars give the Gaussian noise as described in the text. \label{fig-fluct}}
\end{figure}

\begin{figure}
\plottwo{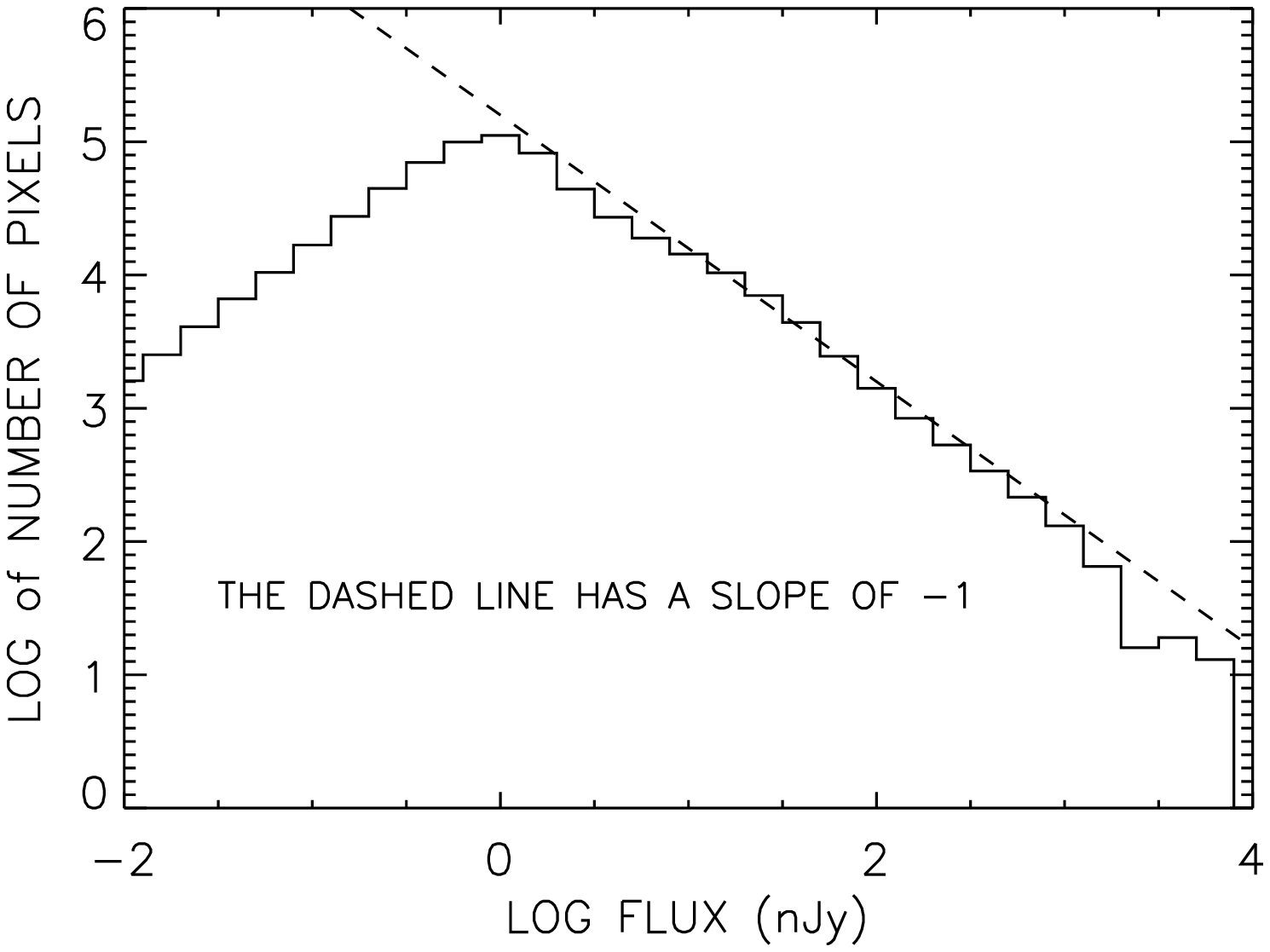}{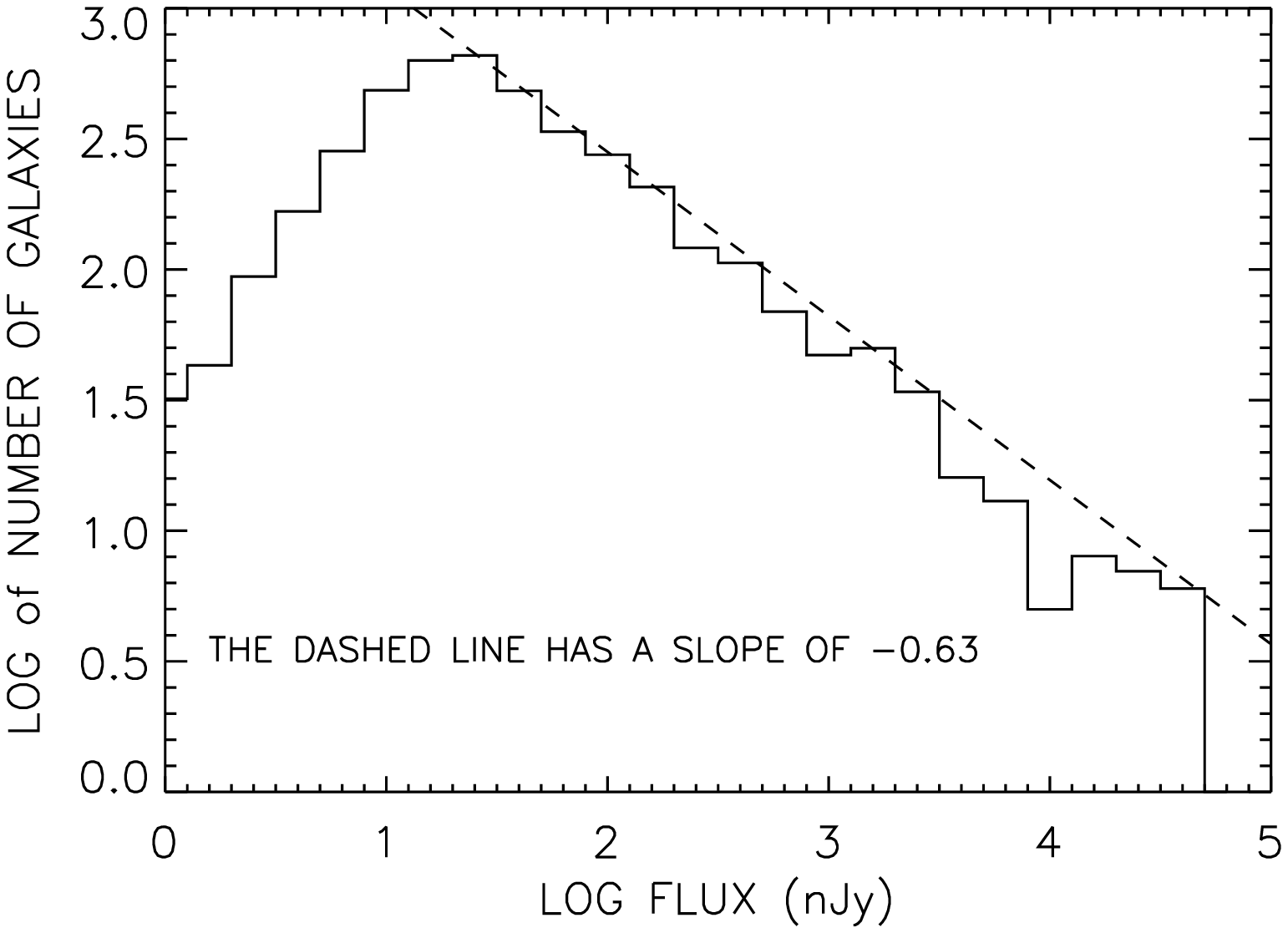}
\caption{a) The left hand figure shows the histogram of the number of
pixels in detected sources having a given flux in nJy in the F160W NUDF image. The dashed
line with slope -1 is not a least squares fit to the data. b)
The right hand figure plots the number of sources for a given total
source flux.  The dashed line with slope -0.63 is not a least squares
fit to the data.\label{fig-fp}}
\end{figure}

\begin{figure}
\plotone{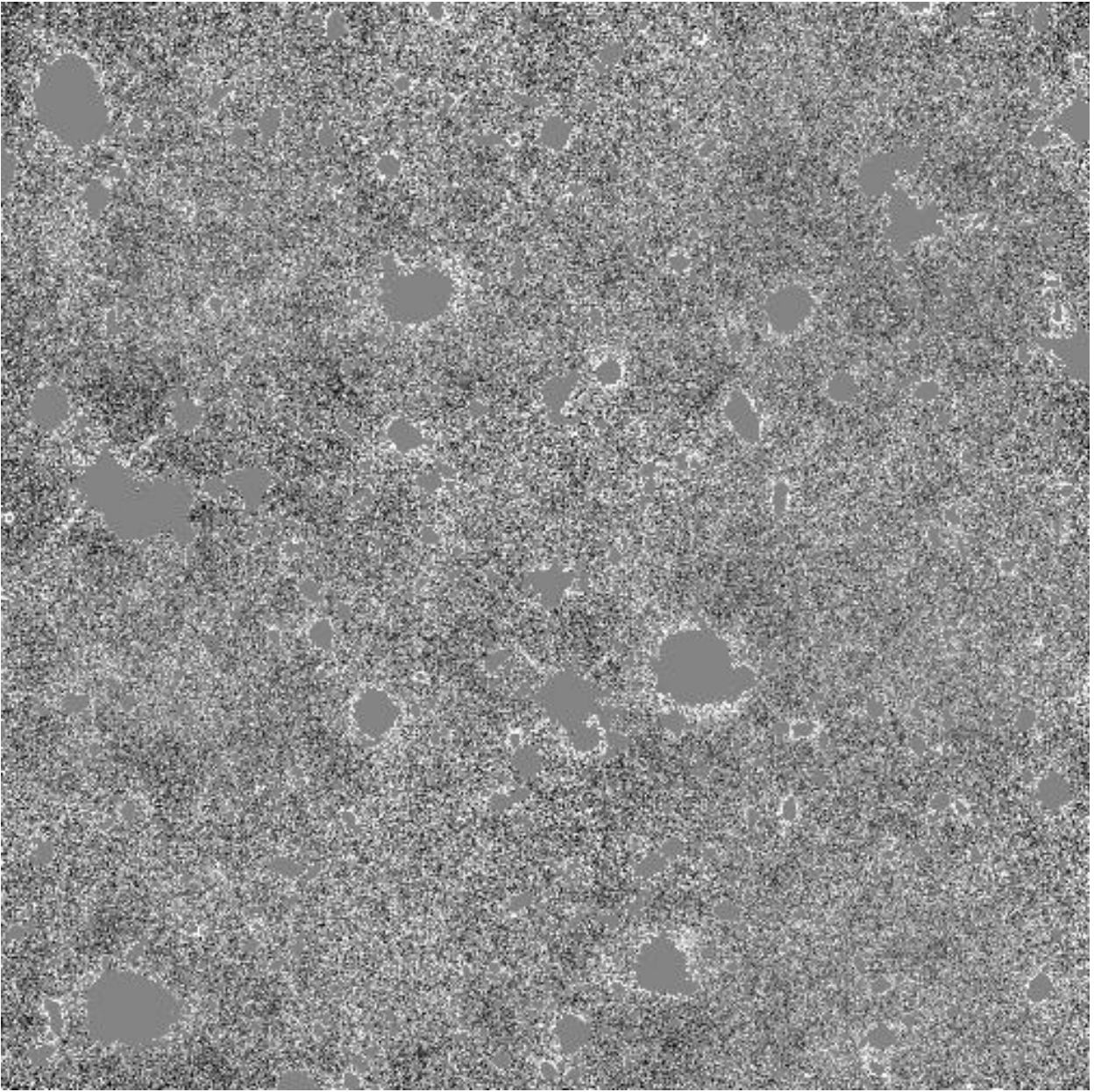}
\caption{A portion of the NUDF F160W image with the flux level in the source 
positions set
to zero.  The image is a linear stretch between $7.5 \times 10^{-4}$ and 
$-7.5 \times 10^{-4}$ ADUs per second which corresponds to $2.9 \times 10^{-12}$
nW m$^{-2}$ sr$^{-1}$.  There is still some residual flux at the edges
of the source removal and from sources below the extraction limit.  At the
scale of this figure it is difficult to see most of the source extractions
which are in general very small. \label{fig-ims}}
\end{figure}

\begin{figure}
\plotone{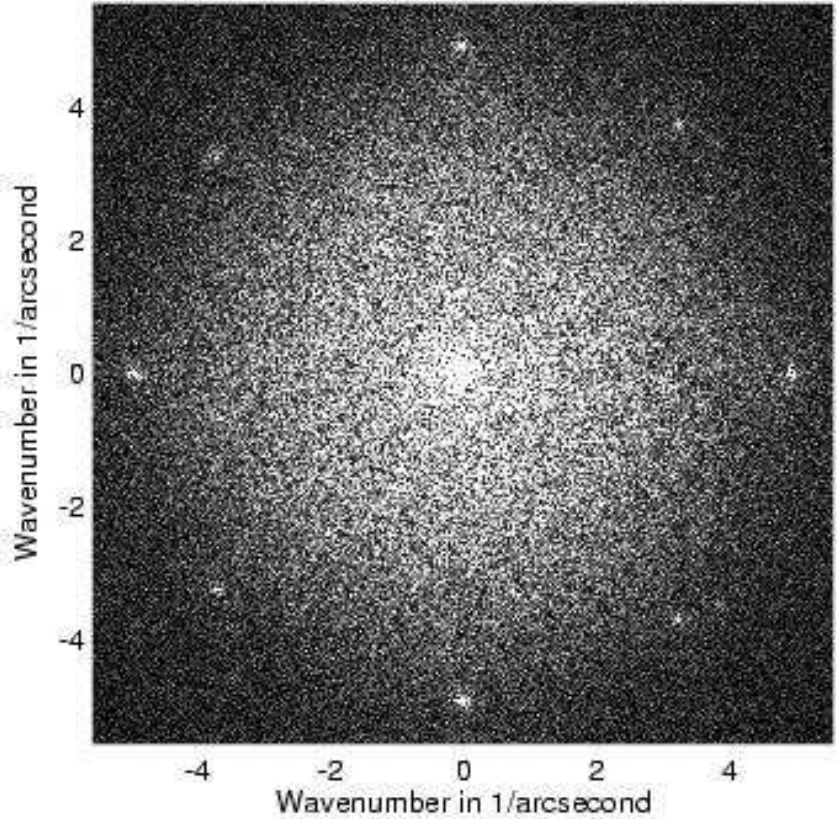}
\caption{The amplitude of the 2 dimensional power spectrum, $\sqrt{f(uv)f^*(u,v)}$ 
in the notation of the
appendix, of the source subtracted image shown in Figure ~\ref{fig-ims}. The
figure is symmetric around the horizontal axis with a linear stretch between 0 and 
1/30 of the maximum value. Other than the white dots in the 
spectrum, there are no artifacts at the spatial scales relevant to this study. 
The white dots are most probably due to the effects of the drizzle procedure as 
discussed in the text. \label{fig-2d}}
\end{figure}

The Fourier series returns a 2 dimensional array of the same size as the image
which is aliased around its midpoint so that we only consider wave numbers in a
semicircle in either the upper or lower half of Figure~\ref{fig-2d}.  We calculate 
the wave number for each
point in the returned array and take the amplitude of the Fourier series at that
point.  We next establish a log spaced vector of $K$ values and find the
\emph{average}, $\mid f_K \mid $, of the $f_K$ values in the bins defined by the $K$
vector to calculate the fluctuation given by the square root of equation~\ref{eq-fl}.
Figure~\ref{fig-fluct} is the plot of those values versus angle where the angle
is $\frac{2\pi}{K}$ and radians are converted to arc seconds.  To assess the influence
of the weight image a similar analysis was done without multiplication by the weight
image.  The only amplitude with a visible change on the scale of Figure~\ref{fig-fluct}
was the very last point.

\clearpage

\begin{deluxetable}{lcccc}
\tabletypesize{\scriptsize}
\tablecaption{Measured Background Fluxes\tablenotemark{a} \label{tab-flx}}
\tablewidth{0pt}
\tablehead{
\colhead{Measurement} & \colhead{Total Flux}   & \colhead{Zodiacal Flux}   &
\colhead{Detected or expected sources\tablenotemark{b}} & \colhead{Residual Flux} 
}
\startdata
NUDF 1.6 $\micron$ &461.9\tablenotemark{c} &455.0\tablenotemark{d} &$6.9^{+3.}_{-0.3}$ &$0.0^{+3.}_{-0.3}$\\
NUDF 1.1 $\micron$ &350.5 &342.20 &$6.3^{+3.}_{-0.3}$ &$0.0^{+3.}_{-0.3}$\\
NHDF 1.6 $\micron$ &327&320 &$7.0^{+3.}_{-0.3}$&$0.0^{+3.}_{-0.3}$\\
NHDF 1.1 $\micron$ &341&334 &$7.0^{+3.}_{-0.3}$&$0.0^{+3.}_{-0.3}$\\
NIRS 1.63 $\micron$ &320.0&224.0&30.1 &65.9\\
NIRS 1.43 $\micron$ &332.1&230.0&32.1 &70.1\\
 \enddata

\tablenotetext{a}{All fluxes are in nW m$^{-2}$ sr$^{-1}$.}
\tablenotetext{b}{The source fluxes for the NUDF and NHDF are for all sources detected
in the images.  The source fluxes for NIRS are the quoted expected fluxes from the
models referenced in \cite{mat05}}
\tablenotetext{c}{The NUDF observations were taken under higher operating temperatures
than the NHDF observations, resulting in a higher F160W background flux.}
\tablenotetext{d}{Includes instrumental background}

\end{deluxetable}

\clearpage

\begin{deluxetable}{lcccccccc}
\tabletypesize{\scriptsize}
\tablecaption{Number of galaxies in each redshift bin \label{tab-nz}}
\tablewidth{0pt}
\tablehead{
\colhead{Redshift Range} & \colhead{0.0-0.5} & \colhead{0.5-1.5} & \colhead{1.5-2.5} & \colhead{2.5-3.5} & \colhead{3.5-4.5} & \colhead{4.5-5.5}& \colhead{5.5-6.5}& \colhead{6.5-10.0}
}
\startdata
Number of galaxies & 511 & 1520 & 931 & 930 & 513 & 218 & 64 & 13\\
Flux in nW m$^{-2}$ sr$^{-1}$ & 2.8 & 3.3 & 0.32 &0.39 & 0.087 &0.014 &0.0046 & 0.00003\\
\enddata

\end{deluxetable}

\clearpage

\clearpage

\clearpage

\clearpage

\clearpage

\clearpage

\begin{figure}
\plotone{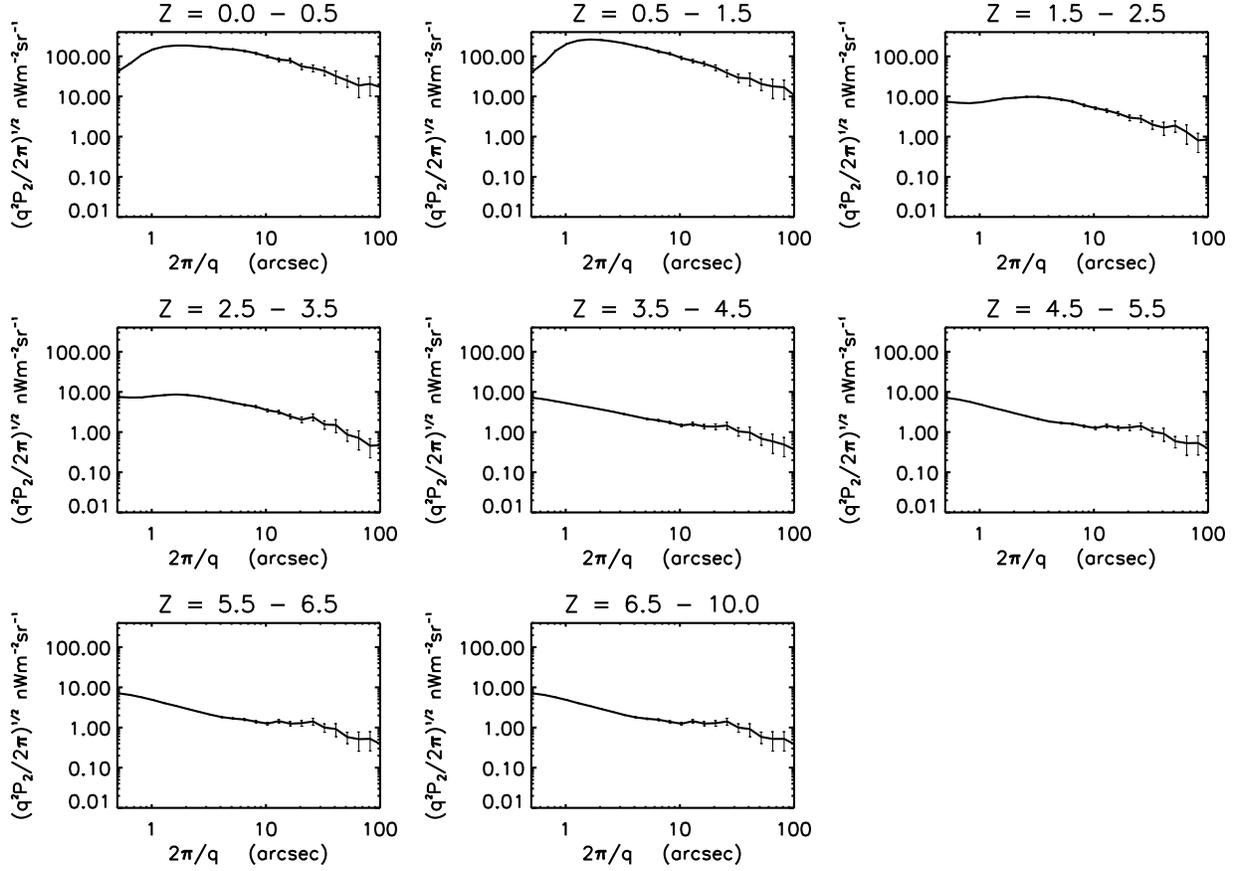}
\caption{The fluctuation spectra with all resolved sources removed except the sources in 
the redshift ranges indicated by the title at the top of each plot. The fluctuation 
spectra for redshifts of 4 and above are almost identical to the spectrum for all 
sources removed shown in more detail in Figure~\ref{fig-fluct}. Any flux not in 
resolved sources is retained in all spectra. \label{fig-fluctz}}
\end{figure}

\clearpage

\begin{figure}
\plottwo{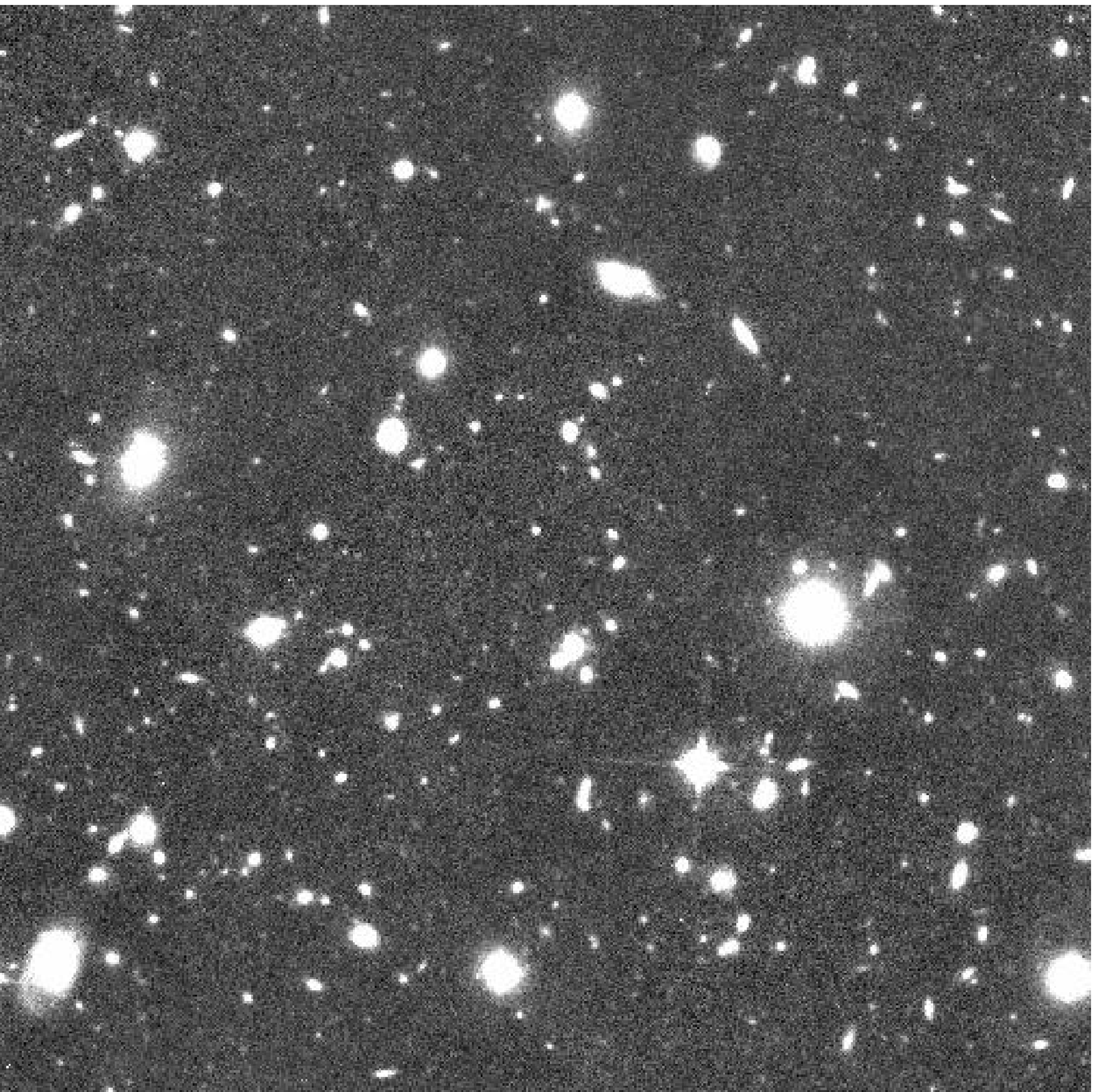}{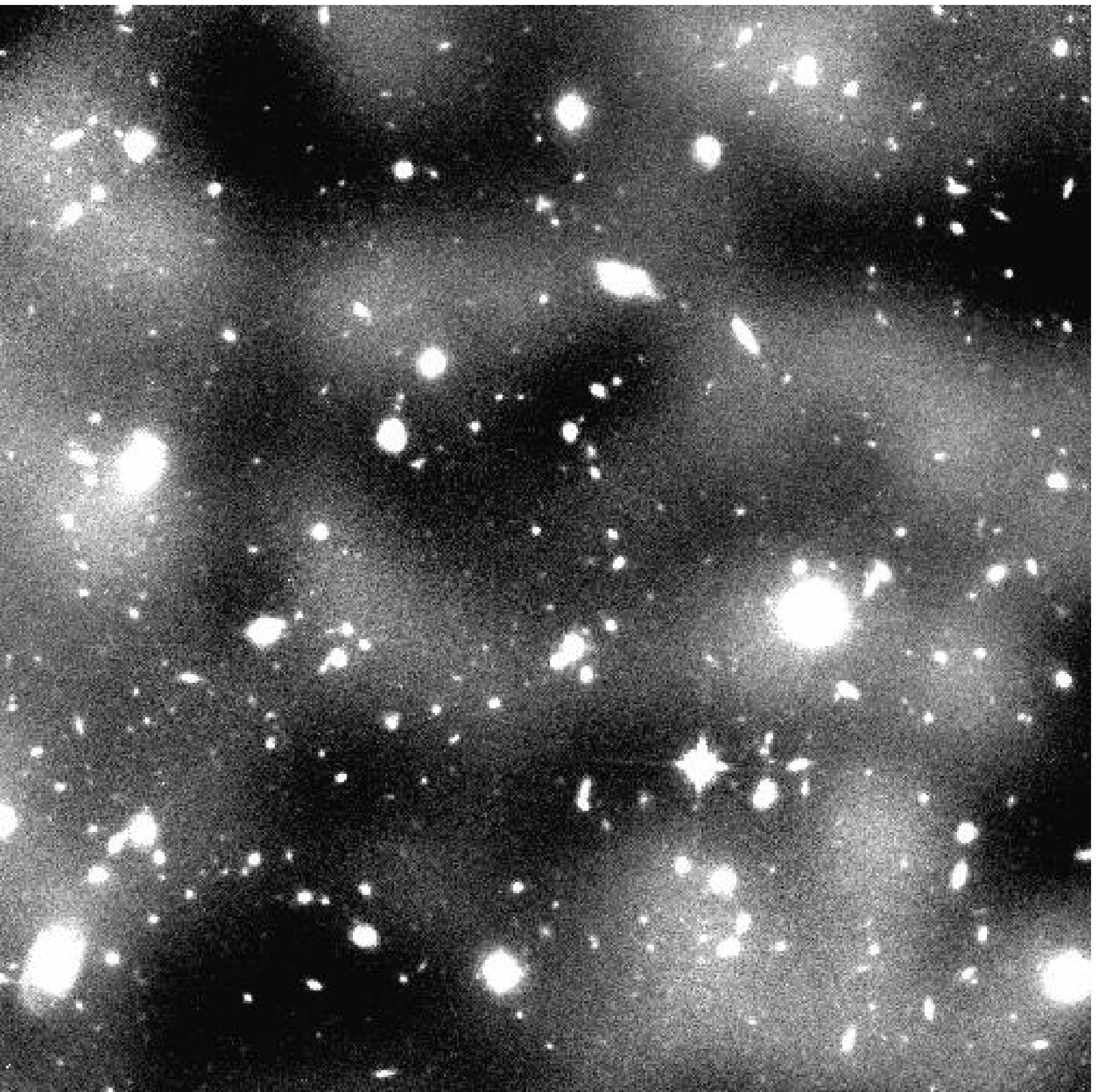}
\caption{a) A portion of the NUDF F160W image without the addition of
a simulated background. b) The same portion of the image with the addition
of diffuse background sources represented by Gaussian images with a FWHM of $10\arcsec$.
All sources have equal flux and the total background flux is 10 times
the flux in detected objects. \label{fig-bac}}
\end{figure}

\end{document}